\documentclass[11pt,american,twoside,a4wide]{article}
\usepackage[T1]{fontenc}
\usepackage[latin1]{inputenc}
\usepackage{latexsym}
\usepackage{amsmath}
\usepackage{bbm}
\usepackage{amssymb}
\usepackage{babel}
\usepackage{amsfonts}
\usepackage{times}
\usepackage{theorem}
\usepackage{epsfig}
\usepackage{enumerate}
\usepackage{color}
\usepackage[active]{srcltx}
\usepackage[colorlinks=true]{hyperref}
\usepackage{fancyhdr}
\newcounter{smallarabics}
\newenvironment{arabicenumerate}
{\begin{list}{{\normalfont\textrm{(\arabic{smallarabics})}}}
  {\usecounter{smallarabics}\setlength{\itemindent}{0cm}
   \setlength{\leftmargin}{5ex}\setlength{\labelwidth}{4ex}
   \setlength{\topsep}{0.75\parsep}\setlength{\partopsep}{0ex}
   \setlength{\itemsep}{0ex}}}
{\end{list}}

\newcounter{smallroman}

\newcommand{\ben}{\begin{arabicenumerate}}  
\newcommand{\een}{\end{arabicenumerate}}  

\newtheorem{theoreme}{Theorem}[section]
\newtheorem{proposition}[theoreme]{Proposition}
\newtheorem{lemma}[theoreme]{Lemma}
\newtheorem{definition}[theoreme]{Definition}
\newtheorem{corollary}[theoreme]{Corollary}

\def\rr{{\mathbb R}}

\def\cc{{\mathbb C}}

\def\textsl{{}}

\def\Im{{\rm Im}\,}

\def\c0inf{C_0^\infty}
\def\bep{\begin{proposition}}
\def\eep{\end{proposition}}

\def\proof{\noindent {\bf Proof.}\ \ }

\def\cH{{\cal  H}}

\def\i{{\rm i}}
\newcommand{\beq}{\begin{equation}}
\newcommand{\eeq}{\end{equation}}
\newcommand{\bear}[1]{\begin{array}{#1}}
\newcommand{\ear}{\end{array}}

\renewcommand{\i}{\mathrm{i}}

\def\qed{$\Box$\medskip}

\def\bel{\begin{lemma}}
\def\eel{\end{lemma}}
\def\bet{\begin{theoreme}}
\def\eet{\end{theoreme}}
\def\bed{\begin{definition}}
\def\eed{\end{definition}}

\def\12{\frac{1}{2}}

\def\Ran{{\rm Ran}\,}

\def\one{{\mathbbm 1}}

\def\cH{{\cal H}}

\begin{document}
\title{The Spectral Structure of the Electronic Black Box Hamiltonian}
\author{P. Grech$^{1,2}$, V. Jak\v{s}i\'c$^{1}$, M. Westrich$^{1}$
\\ \\
$^1$Department of Mathematics and Statistics\\ 
McGill University\\
805 Sherbrooke Street West \\
Montreal,  QC,  H3A 2K6, Canada
\\ \\
$^2$Centre de recherches math\'ematiques\\
Universit\'e de Montr\'eal\\
2920 Chemin de la tour\\
Montr\'eal, QC, H3T 1J4, Canada
}
\maketitle
\thispagestyle{empty}
%%%%%%%%%%%%%%%%%%%%%%%%%%%%%%%%%%%%%%%%%%%%%%%%%%
\begin{quote}
\noindent{\bf Abstract.}
We give results on the absence of singular continuous spectrum of the one-particle Hamiltonian underlying the electronic black box model.
\end{quote}
%%%%%%%%%%%%%%%%%%%%%%%%%%%%%%%%%%%%%%%%%%%%%%%%%%
%%%%%%%%%%%%%%%%%%%%%%%%%%%%%%%%%%%%%%%%%%%%%%%%%

\section{Introduction and Main Results}

We consider a quantum system $S$ with finite dimensional Hilbert
space $\mathcal{H}_{S}$ coupled to a left and a right reservoir with
Hilbert spaces $\mathcal{H}_{\ell}$, $\mathcal{H}_{r}$ respectively.
The Hilbert space of the compound system is given by 
\[
\mathcal{H}:=\mathcal{H_{\ell}}\oplus\mathcal{H}_{S}\oplus\mathcal{H}_{r}\,.
\]
 It carries a free dynamics given by the non-interacting Hamiltonian
\[
H_{0}:=H_{\ell}+H_{S}+H_{r},
\]
 where $H_{j}$ is a self-adjoint operator on
$\mathcal{H}_{j}$, for $j=\ell,S,r$. The coupling between system and reservoirs is modelled as follows.
Let $\chi_{\ell/r}  \in\mathcal{H}_{\ell/r}$,  $\delta_{\ell/r} \in\mathcal{H}_{S}$, 
be distinguished non-zero vectors. The full Hamiltonian is then defined as 
\begin{equation}\label{eq:defH0}
H_{\lambda,\nu}:=H_{0}+\lambda\left[\left(\chi_{\ell},\,\cdot\,\right)\delta_{\ell}+\left(\delta_{\ell},\,\cdot\,\right)\chi_{\ell}\right]+\nu\left[\left(\chi_{r},\,\cdot\,\right)\delta_{r}+\left(\delta_{r},\,\cdot\,\right)\chi_{r}\right]\, ,
\end{equation}
where $\lambda, \nu \in \mathbb{R}$ are control parameters. In the present article we address the question of when the singular continuous spectrum $\sigma_{\text{sc}}(H_{\lambda,\nu})$ is empty.

The fermionic second quantization of the Hamiltonian~(\ref{eq:defH0}) leads to the simplest nontrivial example of the electronic black box model, which has been one of the basic paradigms in the recent developments in non-equilibrium statistical mechanics (see \cite{AJPP1, AJPP2, JKP, N} for references and 
additional information).  The motivation for this article comes from the fact that the absence of singular continuous spectrum is crucial for the rigorous derivation of the Landauer-B\"uttiker formula in transport theory \cite{AJPP2, N}.

It is not hard to see that the cyclic space spanned by $(\chi_{\ell},\chi_{r},\delta_{\ell},\delta_{r})$ and $H_{0}$ agrees with the one spanned by the same vectors and $H_{\lambda,\nu}$. Since $H_{\lambda,\nu}=H_{0}$  on the orthogonal complement of this cyclic space,  for our purposes we may assume without loss of generality that $(\chi_{\ell},\chi_{r},\delta_{\ell},\delta_{r})$ is a cyclic family for $H_{0}$.

Before formulating our main results we gather the following general facts about the connection between boundary values of analytic functions and the spectral measure $\mu^{(\lambda,\nu)}_{\varphi}$ of $H_{\lambda,\nu}$ and $\varphi$ (see, e.g., \cite{J}). In what follows  the quantifier \emph{a.e.} stands for  \emph{almost every}
with respect to Lebesgue measure $\mathcal{L}$ on $\rr$. 

The Borel transform of $\mu_{\varphi}^{(\lambda,\nu)}$,
\[
G_{\lambda,\nu}\left(\varphi,\varphi,z\right):=\int_{\mathbb{R}}\frac{d\mu_{\varphi}^{\left(\lambda,\nu\right)}(E)}{E-z}\,,
\]
 where $z\in\mathbb{C}\backslash\mathbb{R}$ and $\varphi =\chi_{\ell/r},\delta_{\ell/r}$, has the following properties:

\begin{enumerate}
\item The limit 
\[ G_{\lambda, \nu}(\varphi, \varphi, E+\i 0):=\lim_{\epsilon\searrow0}G_{\lambda,\nu}\left(\varphi,\varphi,E+ i\epsilon\right),\]
exists for a.e. $E\in\mathbb{R}$. Moreover, $G_{\lambda, \nu}(\varphi, \varphi, E+\i 0)$ is finite and non-zero for a.e. $E\in\mathbb{R}$.
\item The absolutely continuous part of $\mu_{\varphi}^{(\lambda,\nu)}$ satisfies 
\begin{align*}
d\mu^{(\lambda,\nu)}_{\varphi,\mathrm{ac}}(E)=\frac 1 \pi \mathrm{Im}G_{\lambda,\nu}(\varphi, \varphi, E+i0)dE\, .
\end{align*}

\item The singular part of $\mu^{(\lambda,\nu)}_{\varphi}$ is concentrated on the
set 
\begin{align*}
\{E\in\mathbb{R}\,|\,\lim_{\epsilon\searrow0}{\rm Im}G_{\lambda,\nu}(\varphi,\varphi,E+i\epsilon)=\infty\}\,.
\end{align*}
\end{enumerate}
The connection with spectral theory is established by the spectral theorem through the formula
\begin{align*}
G_{\lambda,\nu}(\varphi,\varphi,z)=(\varphi,\left(H_{\lambda,\nu}-z\right)^{-1}\varphi)\, .
\end{align*}
More generally, we write
\begin{align*}
G_{\lambda,\nu}(\varphi, \psi, z) = \left(\varphi, (H_{\lambda,\nu}-z)^{-1}\psi\right)
\end{align*}
with $\varphi, \psi = \chi_{\ell/r},\delta_{\ell/r}$. 

Throughout this article we shall assume that 
\begin{align} \label{vanish}
G_{0}(\delta_{\ell},\delta_{r},E)= (\delta_\ell, (H_S-E)^{-1}\delta_r)\not \equiv 0 
\end{align}
(double zero indices will be written as a single index, e.g. $G_{0,0}=G_{0}$, $\mu_{\varphi}^{\left(0,0\right)}=\mu_{\varphi}^{(0)}$,
etc.). An equivalent formulation of (\ref{vanish}) is that the cyclic spaces spanned by $\delta_{\ell/r}$ and $H_{S}$  are not orthogonal and hence that the system $S$ does not trivially decouple into two non-interacting subsystems. The case where $G_{0}(\delta_{\ell},\delta_{r},E)\equiv 0$ is simpler and can be treated by the same techniques. However 
this case  has no relevance for applications to non-equilibrium statistical mechanics  that motivated the present work.

To be able to formulate our hypotheses we define 
\[
\mathcal{M}_{0}:=\left\{E\in \mathbb{R}\,|\, G_{0}(\chi_{\ell/r},\chi_{\ell/r}, E+i0)\,\,\text{is finite and non-zero} \right\}.
\]
In view of Property~1, $\mathcal{L}(\mathcal{M}_0^{c})=0$ ($A^{c}$ denotes the complement of a set $A$). We also define
\[
\mathcal{M}_{\ell/r}:=\left\{E\in {\cal M}_0 \,|\, 0<\mathrm{Im}G_{0}(\chi_{\ell/r},\chi_{\ell/r},E+i0)<\infty\right\}.
\]
The set ${\cal M}_\ell \cup {\cal M}_r$ is an  essential support of the absolutely continuous spectrum of $H_{\lambda, \nu}$ 
for all $\lambda, \nu \in {\mathbb R}$ (to avoid confusion we recall that the essential support of a.c. spectrum  is usually defined as an equivalence class of Borel sets  with respect to the 
relation $B_1\sim B_2 \Leftrightarrow {\cal L}((B_1\setminus B_2)\cup (B_2\setminus B_1))=0$).

To fix notation we also recall that the projection-valued measure corresponding to $H_{\lambda,\nu}$ has a unique decomposition into absolutely continuous, pure point, and singular continuous part,
\begin{align*}
 \textbf{1}_{B}\left[H_{\lambda,\nu}\right]=   \textbf{1}_{B}^{\mathrm{ac}}\left[H_{\lambda,\nu}\right]+ \textbf{1}_{B}^{\mathrm{pp}}\left[H_{\lambda,\nu}\right]+  \textbf{1}_{B}^{\mathrm{sc}}\left[H_{\lambda,\nu}\right]\, ,
\end{align*}
where $B\subset \mathbb{R}$ is a Borel set.  Our first result reads as follows:

\bet
\label{thmregular} Let $B\subset \mathbb{R}$ be a Borel set such that  
\[
\left(\mathcal{M}_{\ell}\cup\mathcal{M}_{r}\right)^{c}\cap B
\]
 is countable. Then ${\bf 1}_{B}^{\mathrm{sc}}\left[H_{\lambda,\nu}\right]=0$
for all $\lambda,\nu$.
\eet

Theorem~\ref{thmregular} has a relatively short proof and can be generalized in various ways via an application of the Feshbach method 
(see \cite{BFS, DJ}). In the stated form however it contrasts instructively our main result, which has a considerably more technical proof.
\bet
\label{absreg} Let $B\subset \mathbb{R}$ be a Borel set such that  
\[
\mathcal{L}\left(\left(\mathcal{M}_{\ell}\cup\mathcal{M}_{r}\right)^{c}\cap B\right)=0.
\]
 Then ${\bf 1}_{B}^{\mathrm{sc}}\left[H_{\lambda,\nu}\right]=0$
for a.e. $\lambda,\nu$.
\eet
{\bf Remark.} This result can be viewed as an extension of the Simon-Wolff theorems \cite{SW, J, JKP} to the electronic black box Hamiltonian setting. 
 
Since $H_{\lambda,\nu}-H_{0}$ is compact it follows from Weyl's theorem that $\sigma(H_{\lambda,\nu})\backslash\sigma(H_{0})$ is countable and hence $\textbf{1}^{\mathrm{sc}}_{\sigma(H_{0})^{c}}\left[H_{\lambda,\nu}\right]=0$. Thus we may formulate as a direct consequence of Theorems~\ref{thmregular} and \ref{absreg}: 
\begin{corollary}
\label{cor-abs}
\ben
\item If $\left(\mathcal{M}_{\ell}\cup\mathcal{M}_{r}\right)^{c}\cap \sigma(H_{0})$ is countable, then $\sigma_{\mathrm{sc}}(H_{\lambda,\nu})=\emptyset$ for all $\lambda,\nu$.
\item If $\mathcal{L}(\left(\mathcal{M}_{\ell}\cup\mathcal{M}_{r}\right)^{c}\cap \sigma(H_{0}))=0$, then $\sigma_{\mathrm{sc}}(H_{\lambda,\nu})=\emptyset$ for a.e. $\lambda,\nu$.
\een
\end{corollary}

The key ingredient for the proof of Theorem \ref{absreg} and our principal technical result
concerns  spectral averaging for rank two perturbations. Let $\overline{\mu}_{\varphi}^{\left(\kappa\right)}$ 
be a Borel measure on $\rr$ defined by 
\begin{align*}
\overline{\mu}_{\varphi}^{\left(\kappa\right)}\left(B\right):=\int_{\mathbb{R}}\mu_{\varphi}^{\left(\kappa,\kappa'\right)}\left(B\right)d\kappa',
\end{align*}
where $\kappa=\nu$ and $\kappa'=\lambda$ if $\varphi=\chi_{\ell},\delta_{\ell}$, and
 $\kappa=\lambda$ and $\kappa'=\nu$ if $\varphi=\chi_{r},\delta_{r}$. 
\bet[Spectral averaging]
\label{thm:spectralaveraging1} There is a finite set $\mathcal{N}\subset\mathbb{R}$,
independent of $\lambda,\nu$, such that for a.e. $\kappa$, $\overline{\mu}_{\varphi}^{\left(\kappa\right)}\upharpoonright_{\mathbb{R}\backslash\mathcal{N}}$ is absolutely continuous with respect to ${\cal L}\upharpoonright_{\mathbb{R}\backslash\mathcal{N}}$.

\eet
{\bf Remark 1.} Spectral averaging for rank one perturbations is a classical result  that has been known 
for a long time and refers to the following surprising fact. Given any self-operator $A_{0}$ on a Hilbert space $\mathcal{H}$ and a unit vector $\psi \in \mathcal{H}$, the spectral measure $\mu^{(\lambda)}_{\psi}$ for $A_\lambda:=A_{0}+\lambda(\psi, \cdot) \psi$ averages to the Lebesgue measure: 
\begin{align*}
\overline{\mu}_{\psi}(B)=\int_{\mathbb{R}}\mu^{(\lambda)}_{\psi}(B)d\lambda = \mathcal{L}(B)\, .
\end{align*}
The proof of rank one spectral averaging is simple and  can be found in many places in the literature (see, e.g., \cite{J, S}).\newline
{\bf Remark 2.} 
The set $\mathcal{N}$ need not be empty. 
Let 
\[
\cH_\ell=\cH_r=L^{2}([-2,-1]\cup[1,2], d x)
\]
and $H_\ell=H_r$ be the operator of multiplication by the variable $x$,  $\chi_\ell=\chi_r =\one$
$(\one(x)=1$), and let $\cH_S=\cc$, $H_S=0$, 
$\delta_\ell=\delta_r=\delta=1$. Note that $(\delta, \chi_{\ell}, \chi_{r})$ is a cyclic family for $H_0$ and that Theorem 
\ref{thm:spectralaveraging1} holds for $H_{\lambda, \nu}$.   Let 
\[
\psi_{\lambda,\nu}:=\left(-\frac{\lambda}{x}\right)\oplus 1 \oplus\left(-\frac{\nu}{x}\right).
\]
Then $\psi_{\lambda, \nu}\in \cH$ and $H_{\lambda, \nu}\psi_{\lambda, \nu}=0$ for all $\lambda, \nu$.
The vector  $\psi_{\lambda, \nu}$ is orthogonal to the cyclic subspace spanned by $H_{\lambda, \nu}$ and $(\chi_{\ell}, 
\chi_r)$ and so  $\mu_{\delta}^{(\lambda, \nu)}(\{0\}) >0$ for all $\lambda, \nu$. This implies  
\[
\overline{\mu}_\delta^{(\lambda)}(\{0\})=\overline{\mu}_{\delta}^{(\nu)}(\{0\})>0,
\]
for all $\lambda, \nu$ and  the averaged measures $\overline{\mu}^{(\kappa)}_\delta$ are not absolutely 
continuous with respect to ${\cal L}$. \newline
{\bf Remark 3.} With an additional argument one can show that ${\cal N}\subset \sigma(H_S)$.

As we have already remarked, Theorem \ref{thmregular} can be generalized in many ways by application of standard techniques centered around the 
Feshbach formula. This is not the case with Theorem \ref{absreg}. Our proof is essentially restricted to the simplest example (\ref{eq:defH0})  of 
the electronic black box Hamiltonian and many interesting  questions remain open. 

\bigskip 

{\bf Acknowledgment.} The research of V.J. was partly supported by NSERC.

\section{Proofs}

\subsection{Basic formulas}

\bel
\label{lem:formulas} For $z\in\mathbb{C}\backslash\mathbb{R}$,
\[
G_{\lambda,\nu}(\delta_\ell, \delta_\ell, z)
  =\frac{1}{D(z)}\Big[\Big(1-\nu^{2}G_{0}(\chi_{r},\chi_{r},z)G_{0}(\delta_{r},\delta_{r},z)\Big)G_{0}(\delta_{\ell},\delta_{\ell},z) 
  +\nu^{2}G_{0}(\chi_{r},\chi_{r},z){G_{0}(\delta_{\ell},\delta_{r},z)}\, G_{0}(\delta_{r},\delta_{\ell},z)\Big],
\]
and 
\[
G_{\lambda,\nu} (\chi_{\ell},\chi_{\ell},z)=\frac{1}{D(z)}\left[{G_{0}(\chi_{\ell},\chi_{\ell},z)(1-\nu^{2}G_{0}(\chi_{r},\chi_{r},z)}G_{0}(\delta_{r},\delta_{r},z))\right],\label{crazy2}
\]
where 
\[
\begin{split}
D(z)  =(1&-\nu^{2}G_{0}(\chi_{r},\chi_{r},z)G_{0}(\delta_{r},\delta_{r},z))(1-\lambda^{2}G_{0}(\chi_{\ell},\chi_{\ell},z)G_{0}(\delta_{\ell},\delta_{\ell},z))\\[3mm]
 &-\nu^{2}\lambda^{2}G_{0}(\chi_{r},\chi_{r},z)G_{0}(\chi_{\ell},\chi_{\ell},z) G_{0}(\delta_{\ell},\delta_{r},z)G_{0}(\delta_{r},\delta_{\ell},z)\,.
\end{split}
\]
\eel
\proof The second resolvent identity 
\[
(H_{\lambda,\nu}-z)^{-1}=(H_{0}-z)^{-1}-(H_{\lambda,\nu}-z)^{-1}(H_{\lambda,\nu}-H_{0})(H_{0}-z)^{-1}\label{secresid}
\]
 leads to the system of equations 
\[
\begin{split}
G_{\lambda,\nu}(\delta_{\ell},\delta_{\ell},z)&=  G_{0}(\delta_{\ell},\delta_{\ell},z) -[\lambda G_{\lambda,\nu}(\delta_{\ell},\chi_{\ell},z)G_{0}(\delta_{\ell},\delta_{\ell},z)+\nu G_{\lambda,\nu}(\delta_{\ell},\chi_{r},z)G_{0}(\delta_{r},\delta_{\ell},z)],\\[3mm]
G_{\lambda,\nu}(\delta_{\ell},\chi_{\ell},z)&=  -\lambda G_{\lambda,\nu}(\delta_{\ell},\delta_{\ell},z)G_{0}(\chi_{\ell},\chi_{\ell},z),\\[3mm]
G_{\lambda,\nu}(\delta_{\ell},\chi_{r},z)&=  -\nu G_{\lambda,\nu}(\delta_{\ell},\delta_{r},z)G_{0}(\chi_{r},\chi_{r},z),\\[3mm]
G_{\lambda,\nu}(\delta_{\ell},\delta_{r},z)&=  G_{0}(\delta_{\ell},\delta_{r},z)
  -[\lambda G_{\lambda,\nu}(\delta_{\ell},\chi_{\ell},z)G_{0}(\delta_{\ell},\delta_{r},z)+\nu G_{\lambda,\nu}(\delta_{\ell},\chi_{r},z)G_{0}(\delta_{r},\delta_{r},z)].
\end{split}
\]
Solving the system one derives the formula for  $G_{\lambda, \nu}(\delta_\ell, \delta_\ell, z)$. Similarly, 
\[
\begin{split}
G_{\lambda,\nu}(\chi_{\ell},\chi_{\ell},z)&=  G_{0}(\chi_{\ell},\chi_{\ell},z) -\lambda G_{\lambda,\nu}(\chi_{\ell},\delta_{\ell},z)G_{0}(\chi_{\ell},\chi_{\ell},z),\\[3mm]
G_{\lambda,\nu}(\chi_{\ell},\delta_{\ell},z)&=  -\nu G_{\lambda,\nu}(\chi_{\ell},\chi_{r},z)G_{0}(\delta_{r},\delta_{\ell},z)-\lambda G_{\lambda,\nu}(\chi_{\ell},\chi_{\ell},z)G_{0}(\delta_{\ell},\delta_{\ell},z),\\[3mm]
G_{\lambda,\nu}(\chi_{\ell},\chi_{r},z)&=  -\nu G_{\lambda,\nu}(\chi_{\ell},\delta_{r},z)G_{0}(\chi_{r},\chi_{r},z),\\[3mm]
G_{\lambda,\nu}(\chi_{\ell},\delta_{r},z)&=  -\nu G_{\lambda,\nu}(\chi_{\ell},\chi_{r},z)G_{0}(\delta_{r},\delta_{r},z) -\lambda G_{\lambda,\nu}(\chi_{\ell},\chi_{\ell},z)G_{0}(\delta_{\ell},\delta_{r},z),
\end{split}
\]
 and the formula for $G_{\lambda, \nu}(\chi_\ell, \chi_\ell, z)$ follows. \qed
\vphantom{}

\subsection{Proof of Theorem \ref{thmregular}}
Let 
\begin{align*}
\mathcal{S}:=\{E\in \mathbb{R}\, |\, G_{0}(\delta_{\ell},\delta_{r},E)=0  \}.
\end{align*}
The  Condition~(\ref{vanish}) ensures that ${\cal S}$ is a finite set. Recall also our standing assumption that $(\chi_{\ell},\chi_{r},\delta_{\ell},\delta_{r})$
is a cyclic family for $H_0$ and hence for $H_{\lambda, \nu}$ for all $\lambda, \nu$. Thus, to prove Theorem \ref{thmregular} it suffices to show that for $E\in(\mathcal{M}_{\ell}\cup\mathcal{M}_{r})\backslash (\sigma(H_{S})\cup \mathcal{S})$ and 
$\varphi=\chi_{\ell/r},\delta_{\ell/r}$, the limits 
\begin{equation}
\lim_{\epsilon \searrow 0}\mathrm{Im}G_{\lambda,\nu}(\varphi,\varphi,E+\i \epsilon)
\label{sick}
\end{equation}
exist and are finite for all $\lambda,\nu$. 

If $\lambda=\nu=0$ there is nothing to prove and hence we may assume
that at least one of the parameters is non-zero. By symmetry it suffices
to consider the cases  $\varphi=\chi_{\ell},\delta_{\ell}$. Finally, it follows from  the definition of ${\cal M}_{\ell/r}$ and Lemma 
\ref{lem:formulas} that it suffices to show that for $E\in(\mathcal{M}_{\ell}\cup\mathcal{M}_{r})\backslash (\sigma(H_{S})\cup \mathcal{S})$ we have
$D(E)\not=0$, where 
\[
\begin{split}
D(E)  &=
(1-\nu^{2}G_{0}(\chi_{r},\chi_{r},E+i 0)G_{0}(\delta_{r},\delta_{r},E))(1-\lambda^{2}G_{0}(\chi_{\ell},\chi_{\ell},E)G_{0}(\delta_{\ell},\delta_{\ell},E))\\[3mm]
 &-\nu^{2}\lambda^{2}G_{0}(\chi_{r},\chi_{r},E+i 0)G_{0}(\chi_{\ell},\chi_{\ell},E+i 0) G_{0}(\delta_{\ell},\delta_{r},E)G_{0}(\delta_{r},\delta_{\ell},E).
\end{split}
\]
We argue by contradiction.
Suppose that $D(E)=0$ for some $E\in(\mathcal{M}_{\ell}\cup\mathcal{M}_{r})\backslash (\sigma(H_{S})\cup \mathcal{S})$. 
Set $a  =G_{0}(\delta_{\ell},\delta_{\ell},E)$, $b  =G_{0}(\delta_{r},\delta_{r},E)$, 
$d  =G_{0}(\delta_{\ell},\delta_{\ell},E)G_{0}(\delta_{r},\delta_{r},E)-G_{0}(\delta_{\ell},\delta_{r},E)G_0(\delta_\ell, \delta_r, E)$. 
Since
\[G_{0}(\delta_{\ell},\delta_{r},E)G_{0}(\delta_{r},\delta_{\ell},E)=|G_{0}(\delta_{\ell},\delta_{r},E)|^{2},\]
the numbers $a, b, d$ are real. We also set $l  =G_{0}(\chi_{\ell},\chi_{\ell},E+i0)$, 
$r  =G_{0}(\chi_{r},\chi_{r},E+i0)$. Then the  relation  $D(E)=0$ can be written as 
\[
1-\lambda^{2}al-\nu^{2}br+\lambda^{2}\nu^{2}drl=0,
\]
or equivalently, as 
\begin{equation}
1-\nu^{2}br=\lambda^{2}l(a-\nu^{2}dr) \label{intereq}.
\end{equation}
 Multiplying both sides of (\ref{intereq}) with $a-\nu^{2}d\overline{r}$ yields 
\begin{align*}
a-\nu^{2}d\overline{r}-\nu^{2}bar+\nu^{4}bd|r|^{2}=\lambda^{2}l|a-\nu^{2}dr|^{2}.
\end{align*}
Taking imaginary parts we derive 
\[
\nu^{2}d\mathrm{Im}r-\nu^{2}ba\mathrm{Im}r  =\lambda^{2}\mathrm{Im}l|a-\nu^{2}dr|^{2}.\]
The last equation  is equivalent to 
\begin{equation}
-\nu^{2}|G_{0}(\delta_{\ell},  \delta_{r},E)|^{2}\mathrm{Im}G_{0}(\chi_{r},\chi_{r},E+i0) =\lambda^{2}\mathrm{Im}G_{0}(\chi_{\ell},\chi_{\ell},E+i0)|a-\nu^{2}dG_{0}(\chi_{r},\chi_{r},E+i0)|^{2}\,.\label{aux1}
\end{equation}
 By symmetry we obtain in addition 
\begin{equation}
-\lambda^{2}|G_{0}(\delta_{\ell},\delta_{r},E)|^{2}\mathrm{Im}  G_{0}(\chi_{\ell},\chi_{\ell},E+i0)
  =\nu^{2}\mathrm{Im}G_{0}(\chi_{r},\chi_{r},E+i0)|b-\lambda^{2}dG_{0}(\chi_{\ell},\chi_{\ell},E+i0)|^{2}\,.\label{aux2}
\end{equation}
The right hand sides of Equations (\ref{aux1}, \ref{aux2}) are
non-negative, whereas at least one of the left hand sides is strictly
negative which is a contradiction. \qed
\vphantom{}

\subsection{Proof of Theorem \ref{absreg}}

It suffices to show that, for a.e. $\lambda,\nu$, $\textbf{1}_{ A}\left[H_{\lambda,\nu}\right]\varphi=0$ for the Lebesgue zero set $A:=( \mathcal{M}_{\ell}\cup\mathcal{M}_{r} )^{c} \cap B \backslash\mathcal{N}$ and $\varphi=\delta_{\ell/r}, \chi_{\ell/r}$. 
This however follows from Theorem \ref{thm:spectralaveraging1} since  for
a.e. $\kappa$ ($\kappa=\nu$ and $\kappa'=\lambda$ if $\varphi=\chi_{\ell},\delta_{\ell}$;
 $\kappa=\lambda$ and $\kappa'=\nu$ if $\varphi=\chi_{r},\delta_{r}$)
we have 
\begin{align*}
0=\overline{\mu}_{\varphi}^{\left(\kappa\right)}(A)=\intop_{\mathbb{R}}\mu_{\varphi}^{(\kappa,\kappa')}(A)d\kappa'= 
\intop_{\mathbb{R}}\|\textbf{1}_{ A}\left[H_{\kappa,\kappa^\prime}\right]\varphi\|^2d\kappa'.
\end{align*}

\subsection{Proof of Theorem \ref{thm:spectralaveraging1}}
Throughout the proof we shall omit  standard  measurability arguments (they can be found, for example,  in the lecture notes \cite{J}).  

We start with some preliminaries. By the symmetry $\ell\leftrightarrow r$ it is sufficient to consider
the case $\kappa=\nu$, $\varphi=\chi_{\ell},\delta_{\ell}$. The singular part of $\mu_{\varphi}^{\left(\lambda,\nu\right)}$, 
denoted $\mu_{\varphi, {\rm sing}}^{\left(\lambda,\nu\right)}$,
is concentrated on the set
\begin{align*}
S_{\varphi}(H_{\lambda,\nu}):=\left\{E\in\mathbb{R}\,\big|\,\lim_{\epsilon \searrow 0}\mathrm{Im}G_{\lambda,\nu}(\varphi,\varphi,E+i\epsilon)=\infty\right\}.
\end{align*}
The Poisson transform of the averaged spectral measure $\overline{\mu}_{\varphi}^{\left(\nu\right)}$ can be computed by the residue theorem,
\begin{eqnarray}
\nonumber
P_{\varphi}^{\left(\nu\right)}\left(E+i\epsilon\right)	&:=&\int_{\mathbb{R}}	\mathrm{Im}\,G_{\lambda,\nu}\left(\varphi,\varphi,E+i\epsilon\right)d\lambda \\
\nonumber
	&=&	\int_{\mathbb{R}}	\mathrm{Im}\, \left[\frac{G_{0,\nu}\left(\varphi,\varphi,E+i\epsilon\right)}{1-\lambda^{2}G_{0,\nu}\left(\varphi,\varphi,E+i\epsilon\right)G_{0,\nu}\left(\psi,\psi,E+i\epsilon\right)}\right]d\lambda \\
	&=&	\mathrm{Im}\left(i\pi \sqrt{\frac{G_{0,\nu}\left(\varphi,\varphi,E+i\epsilon\right)}{G_{0,\nu}\left(\psi,\psi,E+i\epsilon\right)}}\right)\geq0,
\label{eq:Poissontrafo1}
\end{eqnarray}
where the last inequality in (\ref{eq:Poissontrafo1}) fixes the branch of the square root and $\varphi, \psi 
\in \{\chi_\ell, \delta_{\ell}\}$, $\varphi\not=\psi$. The  singular part
 $\overline{\mu}_{\varphi, {\rm sing}}^{\left(\nu\right)}$ of the averaged spectral measure 
 is concentrated on the set 
\begin{align}\label{averagedS}
\overline{S}_{\varphi}^{\left(\nu\right)}:=\left\{ E\in\mathbb{R}\,\big|\,\lim_{\epsilon \searrow 0}P_{\varphi}^{\left(\nu\right)}\left(E+i\epsilon\right)=\infty\right\}.
\end{align}
Since 
\[
\overline{\mu}_{\varphi, {\rm sing}}^{\left(\nu\right)}\leq \int_\rr \mu_{\varphi, {\rm sing}}^{\left(\lambda,\nu\right)}d\lambda,
\]
we have 
\begin{equation}
\overline{S}_{\varphi}^{\left(\nu\right)}\subset \bigcup_{\lambda\in\mathbb{R}\backslash\left\{ 0\right\} }S_{\varphi}\left(H_{\lambda,\nu}\right).
\label{fri-sick}
\end{equation}
We also introduce the sets
\begin{equation}
C_{\varphi,\psi}^{\left(\nu\right)}:=\left\{ E\in\mathbb{R}\,\big|\,
\lim_{\epsilon \searrow 0}\left|G_{0,\nu}\left(\varphi,\varphi,E+i\epsilon\right)\right|=\infty, \,\,
\lim_{\epsilon \searrow  0}G_{0,\nu}\left(\psi,\psi,E+i\epsilon \right)=0\right\}\label{eq:Cvarphipsinu}
\end{equation}
for $\varphi,\psi\in\left\{ \chi_{\ell},\delta_{\ell}\right\} $, $\varphi\neq\psi$. Clearly,  ${\cal L}(C_{\varphi,\psi}^{\left(\nu\right)})=0$.
\bel
\[
\overline{\mathcal{S}}_{\varphi}^{\left(\nu\right)} \subset C_{\varphi,\psi}^{\left(\nu\right)}.
\]
\label{cats}
\eel
{\bf Remark.} Since $G_{0, \nu}(\chi_\ell, \chi_\ell, E+i \epsilon)= G_0(\chi_\ell, \chi_\ell, E+i \epsilon)$, we have in 
particular 
\[
\begin{split}
\overline{\mathcal{S}}_{\chi_\ell}^{\left(\nu\right)}&\subset \left\{ E\in\mathbb{R}\,\big|\,
\lim_{\epsilon \searrow 0}\left|G_{0}\left(\chi_\ell,\chi_\ell, E+i\epsilon\right)\right|=\infty\right\},\\[3mm]
\overline{\mathcal{S}}_{\delta_\ell}^{\left(\nu\right)}&\subset \left\{ E\in\mathbb{R}\,\big|\,
\lim_{\epsilon \searrow 0}G_{0}\left(\chi_\ell,\chi_\ell, \varphi,E+i\epsilon\right)=0\right\}.
\end{split}
\]
\proof
Let $E\in\overline{\mathcal{S}}_{\varphi}^{\left(\nu\right)}$. Set 
\[
v\left(\epsilon\right):=G_{0,\nu}\left(\varphi,\varphi,E+i\epsilon\right),\quad w\left(\epsilon\right):=G_{0,\nu}\left(\psi,\psi,E+i\epsilon\right),
\]
and note that 
\[
G_{\lambda,\nu}\left(\varphi,\varphi,E+i\epsilon\right)=\frac{ v(\epsilon)}{1-\lambda^2 v(\epsilon)w(\epsilon)}.
\]
By (\ref{fri-sick}), $E\in \mathcal{S}_{\varphi}\left(H_{\lambda,\nu}\right)$ 
for some $\lambda \not=0$ and  $\lim_{\epsilon\searrow0}\mathrm{Im}G_{\lambda,\nu}\left(\varphi,\varphi,E+i\epsilon\right)=\infty$
is equivalent to 
\begin{equation*}
\lim_{\epsilon\searrow0}\mathrm{Im}\frac{1}{\frac{1}{v\left(\epsilon\right)}-\lambda^{2}w\left(\epsilon\right)}=\infty.\label{eq:Imsingsupp}
\end{equation*}
 Since for any $z\in\mathbb{C}$, 
\begin{equation*}
\left|\mathrm{Im}\frac{1}{z}\right|=\left|\frac{-\mathrm{Im}z}{\left|z\right|^{2}}\right|\le\frac{1}{\left|\mathrm{Im}z\right|},\label{eq:Imz}
\end{equation*}
 we observe that  $\left|\mathrm{Im}z^{-1}\right|\to\infty$ implies $\mathrm{Im}z\to0$
and hence also $\mathrm{Re}z\to0$. This leads to
\begin{eqnarray}\label{goes0}
\frac{1}{v\left(\epsilon\right)}-\lambda^2w\left(\epsilon\right)=:h\left(\epsilon\right)\to0,
\end{eqnarray}as $\epsilon\to0$. 

Suppose that  $\sup_n|v\left(\epsilon_n\right)|<\infty$ along some sequence $\epsilon_n \downarrow 0$. Since
\begin{align*}
\frac{v\left(\epsilon\right)}{w\left(\epsilon\right)}=\frac{\lambda^2v\left(\epsilon\right)^2}{1-v\left(\epsilon\right)h\left(\epsilon\right)},
\end{align*} 
we have (recall (\ref{averagedS}))
\[
\infty=\lim_{n\rightarrow \infty}\left|\mathrm{Re}\left(\sqrt{\frac{v\left(\epsilon_n\right)}{w\left(\epsilon_n\right)}}\right)\right|\leq\limsup_{n\rightarrow \infty}\sqrt{\left|\frac{v\left(\epsilon_n\right)}{w\left(\epsilon_n\right)}\right|}=|\lambda| \limsup_{n\rightarrow \infty}\left|v\left(\epsilon_n\right)\right|<\infty,
\] which is a contradiction. Hence   $\lim_{\epsilon\searrow0}\left|v\left(\epsilon\right)\right|=\infty$. This fact and (\ref{goes0}) yield that   
 \begin{align*}
 \lim_{\epsilon \searrow 0} w(\epsilon)=0.
 \end{align*} \qed
\vphantom{}

For $E\in \rr\setminus \sigma(H_S)$ we  set 
\[ d(E)=G_0(\delta_\ell, \delta_\ell, E)G_0(\delta_r, \delta_r, E)-
G_0(\delta_\ell, \delta_r, E)G_0(\delta_r, \delta_\ell, E).
\]
Let 
\[
{\cal N}:= \left\{ E\in \rr\setminus \sigma(H_S)\,\big|\, G_0(\delta_\ell,\delta_\ell,  E)G_0(\delta_r , \delta_r, E)G_0(\delta_\ell, \delta_r, E)d(E)=0\right\} \cup 
\sigma(H_S). 
\]
The set ${\cal N}$ is finite.\bel 
\ben
\item $C_{\varphi, \psi}^{(0)}\subset {\cal N}$. 
\item For $\nu\not=0$, 
\[C_{\chi_\ell, \delta_\ell}^{(\nu)}\setminus {\cal N}= \left\{ E\in \rr\,\big|\, 
\lim_{\epsilon \searrow 0}|G_0(\chi_\ell, \chi_\ell, E+i \epsilon)|=\infty,\,\,
\lim_{\epsilon \searrow 0}G_0(\chi_r, \chi_r, E+i \epsilon)=\frac{G_0(\delta_\ell, \delta_\ell, E)}{\nu^2 d(E)}\right\}.
\]
\item For $\nu\not=0$, 
\[C_{\delta_\ell, \chi_\ell}^{(\nu)}\setminus {\cal N}= \left\{ E\in \rr\,\big|\, 
\lim_{\epsilon \searrow 0}|G_0(\chi_\ell, \chi_\ell, E+i \epsilon)|=0,\,\,
\lim_{\epsilon \searrow 0}G_0(\chi_r, \chi_r, E+i \epsilon)=\frac{1}{\nu^2G_0(\delta_r, \delta_r, E)}\right\}.
\]
\een
\label{no-end}
\eel
\proof We will deal with the case $\varphi=\chi_\ell$, $\psi=\delta_\ell$, the other case is similar. Note that 
\[ G_{0, \nu}(\chi_\ell, \chi_\ell, E+i \epsilon)= G_0(\chi_\ell, \chi_\ell, E+i \epsilon),
\]
and that by   Lemma \ref{lem:formulas},
\[
G_{0,\nu}(\delta_{\ell},\delta_{\ell},  E+i\epsilon)
=  G_{0}\left(\delta_{\ell},\delta_{\ell},E+i\epsilon\right)
  +\nu^{2}\frac{G_{0}\left(\chi_{r},\chi_{r},E+i\epsilon\right)G_{0}\left(\delta_{\ell},\delta_{r},E+i\epsilon\right)G_{0}\left(\delta_{r},\delta_{\ell},E+i\epsilon\right)}{1-\nu^{2}G_{0}\left(\chi_{r},\chi_{r},E+i\epsilon\right)G_{0}\left(\delta_{r},\delta_{r},E+i\epsilon\right)}.\label{eq:Fnudeltaell}
\]
Part (1) is now obvious and  simple algebra yields Part (2). \qed

We are now ready to complete the proof of Theorem \ref{thm:spectralaveraging1}. We shall deal with the case
$\varphi=\chi_\ell$. The argument is identical in the case  $\varphi=\delta_\ell$. 

Let
\[
\Omega_{\nu}:=C_{\chi_{\ell},\delta_{\ell}}^{\left(\nu\right)}\backslash\mathcal{N}.
\]
By Lemma \ref{cats},  the singular part of $\overline{\mu}_{\chi_\ell}^{\left(\nu\right)}\restriction_{\mathbb{R}\backslash\mathcal{N}}$,
is concentrated on $\Omega_\nu$.  Our strategy is to show that 
\begin{equation}
\textbf{ 1}_{\Omega_\nu}[H_{\lambda, \nu}]=0
\label{stra}
\end{equation}
for a.e. $\lambda, \nu$. 
This implies the statement since 
\[
\overline{\mu}_{\chi_\ell}^{\left(\nu\right)}(\Omega_\nu)=
\int_\rr \|\textbf{1}_{\Omega_\nu}[H_{\lambda, \nu}]\chi_\ell\|^2d\lambda.
\]
Let 
\[
{\cal A}:=\{ E\in \rr\,\big|\, \lim_{\epsilon \searrow 0}|G_0(\chi_\ell, \chi_\ell, E+i \epsilon)|=\infty, \,\, \lim_{\epsilon \searrow 0}G_0(\chi_r, \chi_r, E+i\epsilon) \,\,\hbox{exists and is finite and non-zero}\}.
\]
${\cal L}({\cal A})=0$ and by Lemma \ref{no-end}, $\Omega_\nu \subset {\cal A}$ for all $\nu$.  We claim that for  all $\lambda$,  
\begin{equation}
\begin{split}
\overline{\mu}_{\chi_{r}}^{\left(\lambda\right)}({\cal A})
&=\int_{\rr}\mu_{\chi_r}^{(\lambda, \nu)}({\cal A})d\nu =0,\\[3mm]
\overline{\mu}_{\delta_{r}}^{\left(\lambda\right)}({\cal A})&
=\int_{\rr}\mu_{\delta_r}^{(\lambda, \nu)}({\cal A})d\nu =0.
\end{split}
\label{eq:averagedmeasuresac}
\end{equation}
To establish these relations, define $C_{\varphi,\psi}^{\left(\lambda\right)}$ by (\ref{eq:Cvarphipsinu})
with $\lambda$ replacing $\nu$ and $\varphi,\psi\in\left\{ \chi_{r},\delta_{r}\right\} $,
$\varphi\neq\psi$.   By Lemma \ref{cats} and the remark after it, the singular parts of the measures 
$\overline{\mu}_{\chi_{r}}^{\left(\lambda\right)}$, $\overline{\mu}_{\delta_{r}}^{\left(\lambda\right)}$ are concentrated 
on the set 
\[ {\cal A}^\prime=\left\{ E\in \rr\,\big|\, \lim_{\epsilon \searrow 0}|G_{0}(\chi_{r},\chi_{r},E+i\epsilon)|=\infty\right\} 
\cup \left\{ E\in \rr\,\big|\, \lim_{\epsilon \searrow 0}G_{0}(\chi_{r},\chi_{r},E+i\epsilon)=0\right\}.
\]
Since   ${\cal A}\cap {\cal A}^\prime=\emptyset$, the relations  (\ref{eq:averagedmeasuresac}) follow.
As a consequence of (\ref{eq:averagedmeasuresac}), for a given $\lambda$ we have 
\[\textbf{1}_{\Omega_\nu}[H_{\lambda, \nu}]\delta_r=\textbf{1}_{\Omega_\nu}[H_{\lambda, \nu}]\chi_r=0\qquad \hbox{for a.e. $\nu$}.\]
It follows  that for a.e. $\nu$ 
\[
H_{\lambda,\nu}\psi= H_{\lambda, 0}\psi
\]
where $\psi\in\Ran \textbf{1}_{\Omega_{\nu}\cap [-M,M]}\left[H_{\lambda,\nu}\right]$ and  $M>0$. An application 
of the functional calculus gives that for  a.e. $\nu$ and for any bounded Borel function $f$, 
\[ f(H_{\lambda, \nu})\upharpoonright\Ran \textbf{1}_{\Omega_{\nu}}\left[H_{\lambda,\nu}\right]=
 f(H_{\lambda, 0})\upharpoonright\Ran \textbf{1}_{\Omega_{\nu}}\left[H_{\lambda,\nu}\right].
 \]
In particular, for a.e. $\nu$ and all $\psi\in \Ran \textbf{1}_{\Omega_{\nu}}\left[H_{\lambda,\nu}\right]$,
\[\psi ={\bf 1}_{\Omega_\nu}(H_{\lambda, 0})\psi.\] 
Note that  
\[ H_{\lambda, 0}= h_\lambda \oplus H_r,
\]
where 
\[
h_\lambda:=H_{\ell} + H_S + \lambda\left[\left(\chi_{\ell},\,\cdot\,\right)\delta_{\ell}+\left(\delta_{\ell},\,\cdot\,\right)\chi_{\ell}\right]
\]
acts on $\cH_\ell \oplus \cH_S$.  Lemma \ref{no-end} (1) implies that  for all $\nu$ 
\[
\int_\rr \mu_{\chi_\ell}^{(\lambda, 0)}(\Omega_\nu)d \lambda=\int_\rr \mu_{\delta_\ell}^{(\lambda, 0)}(\Omega_\nu)d \lambda=0.
\]
Since $h_\lambda=H_S$ on the orthogonal complement (in $\cH_\ell \oplus \cH_S$) of the cyclic space spanned by $h_\lambda$ and 
$(\chi_\ell, \delta_\ell)$, we derive that 
\[
\textbf{1}_{\Omega_\nu}(h_\lambda)=0\qquad \hbox{for a.e. $\lambda$}.
\]
The singular part of the spectral measure  of $H_r$ and $\chi_r$ is concentrated on the set 
\[\left\{ E\in \rr\,\big|\, \lim_{\epsilon \searrow 0}\Im G_0(\chi_r, \chi_r, E+i \epsilon)=\infty\right\},
\]
and so $\textbf{1}_{\Omega_\nu}(H_r)=0$. Hence, for a.e $\lambda, \nu$ and any $\psi\in \Ran \textbf{1}_{\Omega_{\nu}}\left[H_{\lambda,\nu}\right]$, 
\[
\psi=\textbf{1}_{\Omega_\nu}(H_{\lambda, 0})\psi=0,
\]
and (\ref{stra}) follows. 

\end{document}